\documentclass[ ,final]{aipproc}

\layoutstyle{6x9}

\begin{document}

\title{New Results on Nucleon Resonance Transition Form Factors}


\author{Volker D. Burkert, Inna Aznauryan\thanks{now at Yerevan Physics Institute, Yerevan, Armenia}, and the CLAS Collaboration} {address={Jefferson Lab, 12000 Jefferson Avenue, 
Newport News, Virginia }}

\begin{abstract}
Recent measurements with CLAS at Jefferson Lab of nucleon resonance transition form factors 
for several lower mass states are discussed.
\end{abstract}

\maketitle

\section{Introduction}
\label{intro}

Electroexcitation of nucleon resonances has long been recognized as a 
sensitive tool in exploring the complex nucleon structure at 
varying distances scales. Mapping out the transition helicity amplitudes will tell 
us a great deal about the underlying quark or hadronic structure.   
Most of the recent data have been taken with the CLAS detector~\cite{clas} using the 6 GeV 
polarized electron beam at Jefferson lab. This allows to measure simultaneously 
the entire resonance mass region and a wide range in the photon virtuality $Q^2$.
Several final states are measured simultaneously~\cite{burkert-lee}. 
In this talk I discuss recent results on
the extraction of transition amplitudes for several well-known states 
from pion electroproduction. 

\section{The $N\Delta(1232)$ transition}
\label{sec:ndelta}
\begin{figure}
\includegraphics[width=7cm]{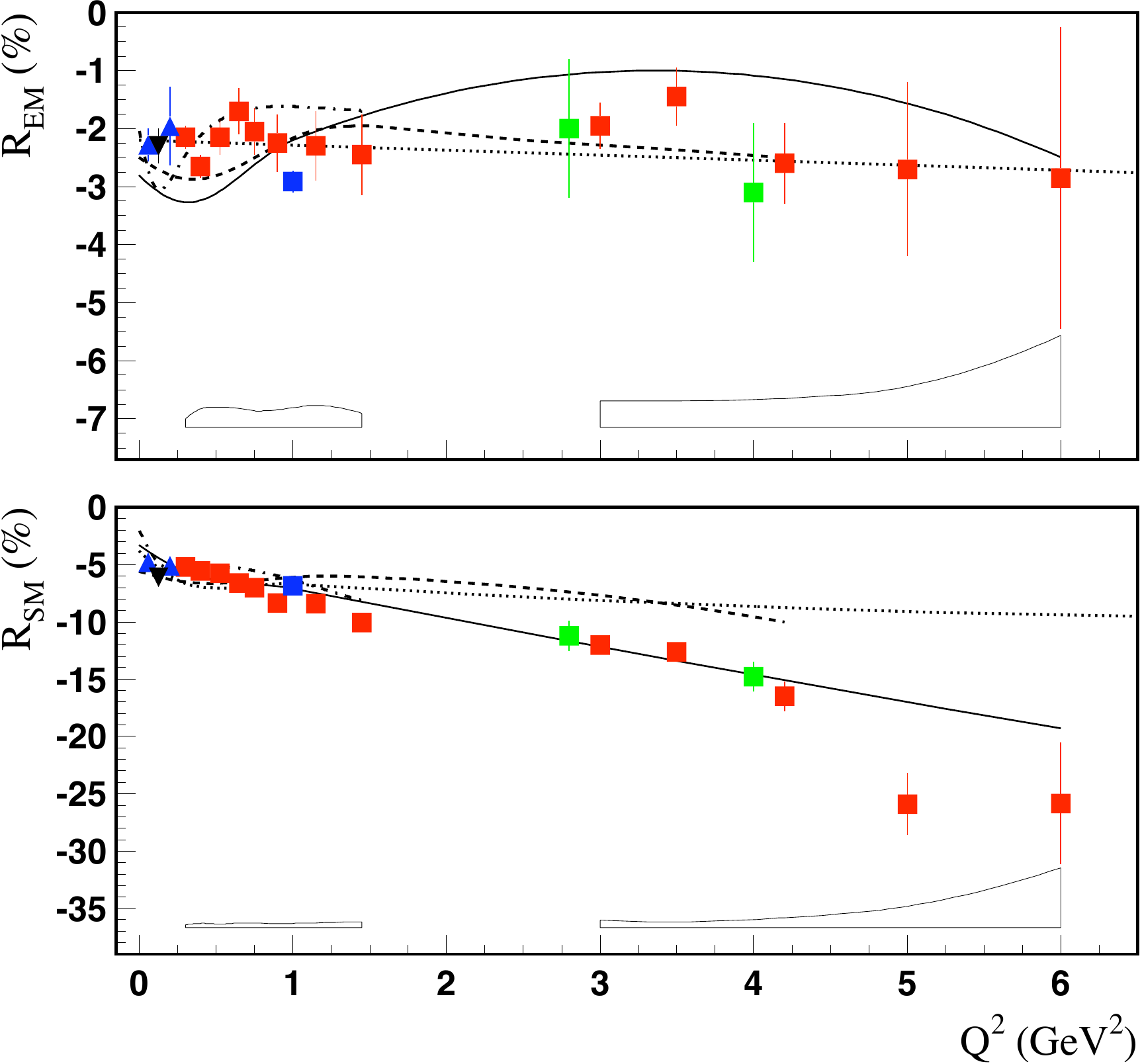}
\caption{$R_{EM}$ and $R_{SM}$ from $p\pi^\circ$ electroproduction.
 $p(e,e^\prime p)\pi^0$. Data from \cite{mami-delta,bates-delta,frolov,kjoo,ungaro,kelly}. }
 \label{remrsm}
\end{figure}

An interesting aspect of nucleon structure at low energies 
is a possible quadrupole deformation of the lowest excited state, the 
$\Delta(1232)$. 
Such a deformation would be evident in finite values of the quadrupole transition 
amplitude $E_{1+}$ and $S_{1+}$, which otherwise would be equal to zero \cite{buchmann}. 
Quadrupole ratios $R_{EM}=E_{1+}/M_{1+}$ and $R_{SM}=S_{1+}/M_{1+}$  are shown
 in Fig.\ref{remrsm}. 
The development of sophisticated phenomenological analysis methods~\cite{maid,janr} over 
the past decade resulted in a consistent picture for these quantities. $R_{EM}$ remains negative, small 
and nearly constant in the entire range $0< Q^2<6$~GeV$^2$. There are no indications that leading 
pQCD contributions are important, which would require  $R_{EM} \rightarrow +1$ ~\cite{carlson}.
 The longitudinal quadrupole ratio $R_{SM}$ also remains negative, but its magnitude 
 rises strongly with increasing $Q^2$.
Simultaneous description of both $R_{EM}$ and $R_{SM}$ is achieved with dynamical 
models that include pion-nucleon interactions explicitly, supporting the idea that most of the 
quadrupole strength in the $N\Delta(1232)$ transition is due to meson effect~\cite{sato,yang} . 
\section{The Roper resonance - one puzzle resolved }
The standard constituent quark model which describes this state 
as a radial excitation of the nucleon, has difficulties to describe basic
features such as the mass, photocouplings, and $Q^2$ evolution. This has 
prompted the development of alternative models involving gluon 
fields~\cite{libuli}, or meson-baryon degrees of freedom~\cite{cano,krewald}.  
\begin{figure}[b]
\includegraphics[width=12cm,height=6cm]{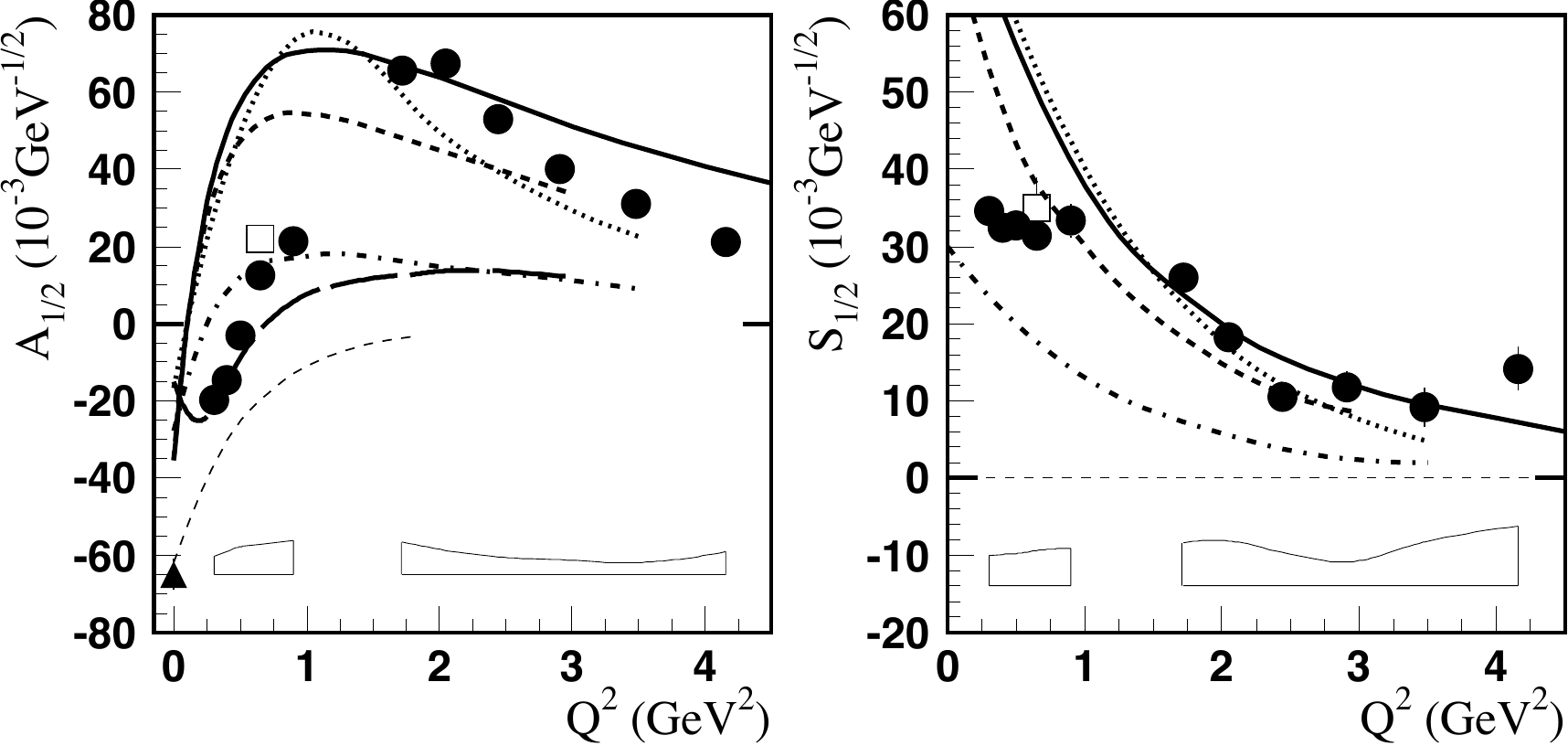}
\caption {Transverse electrocoupling amplitude for the 
Roper $P_{11}(1440)$ (left panel). The full circles 
are the new CLAS results. The squares are previously published results of fits to CLAS data
at low $Q^2$. The right panel shows the longitudinal amplitude. The bold curves are all 
relativistic light front quark model calculations \protect\cite{aznauryan-qm}. The thin dashed 
line is for a gluonic excitation\protect\cite{libuli}. } 
\label{roper}
\end{figure}
Given these different theoretical concept for the structure of the state, 
the question ``what is the nature of the Roper state?'' has been a focus of 
the $N^*$ program with CLAS.  The state is very wide, and pion electroproduction 
data covering a large range in the 
invariant mass W with full center-of-mass angular coverage are key in extracting the 
transition form factors. As an isospin  $I = {1\over 2}$ state, 
the $P_{11}(1440)$ couples more strongly to n$\pi^+$ than to p$\pi^\circ$. Also
contributions of the high energy tail of the $\Delta(1232)$ are much reduced in that 
channel due to the $I = {3\over 2}$ nature of the $\Delta(1232)$. Previous 
studies~\cite{maid07} have mostly used the $p\pi^0$ final state often resulting from 
measurements that focussed on the $\Delta(1232)$ mass region. This analysis 
included new high statistics $n\pi^+$ data that covered the entire mass region 
up to $W=1.7$~GeV.

A large sample of differential cross sections and polarization asymmetry 
data ~\cite{kjoo,kjoo2003,kjoo2004,kjoo2005,hovanes,biselli2008,park08} 
from  CLAS have
been analyzed using the fixed-t dispersion relations approach and the unitary isobar 
model. 
The transverse and longitudinal electrocoupling amplitudes $A_{1/2}$ and $S_{1/2}$ of 
the transition to the $N(1440)P_{11}$ resonance are extracted from fits~\cite{aznauryan-1,aznauryan-2}
to these data, and are shown in Fig.~\ref{roper}.      

At the real photon point $A_{1/2}$ is negative, rises quickly with $Q^2$,  and changes sign near $Q^2=0.5$~GeV$^2$. At $Q^2=2$GeV$^2$ the 
amplitude has about the same magnitude but opposite
sign as at $Q^2=0$. It slowly falls off at high $Q^2$. This remarkable behavior of a sign 
change with $Q^2$ has not been observed before for any nucleon form factor or transition 
amplitude. The longitudinal coupling 
$S_{1/2}$ is smaller than the transverse one. At high $Q^2$ both 
amplitudes are qualitatively described by the 
light front quark models, which strongly suggests that at short distances the Roper behaves 
indeed as expected from a radial excitation 
of the nucleon. The low $Q^2$ behavior is not well described by the LF quark models and 
all fall short of describing the amplitude at the photon point. This suggests that important
contributions, e.g. meson-baryon interactions describing the large distances behavior, 
are missing. 
\begin{figure}[h]
\includegraphics[width=12cm,height=5cm]{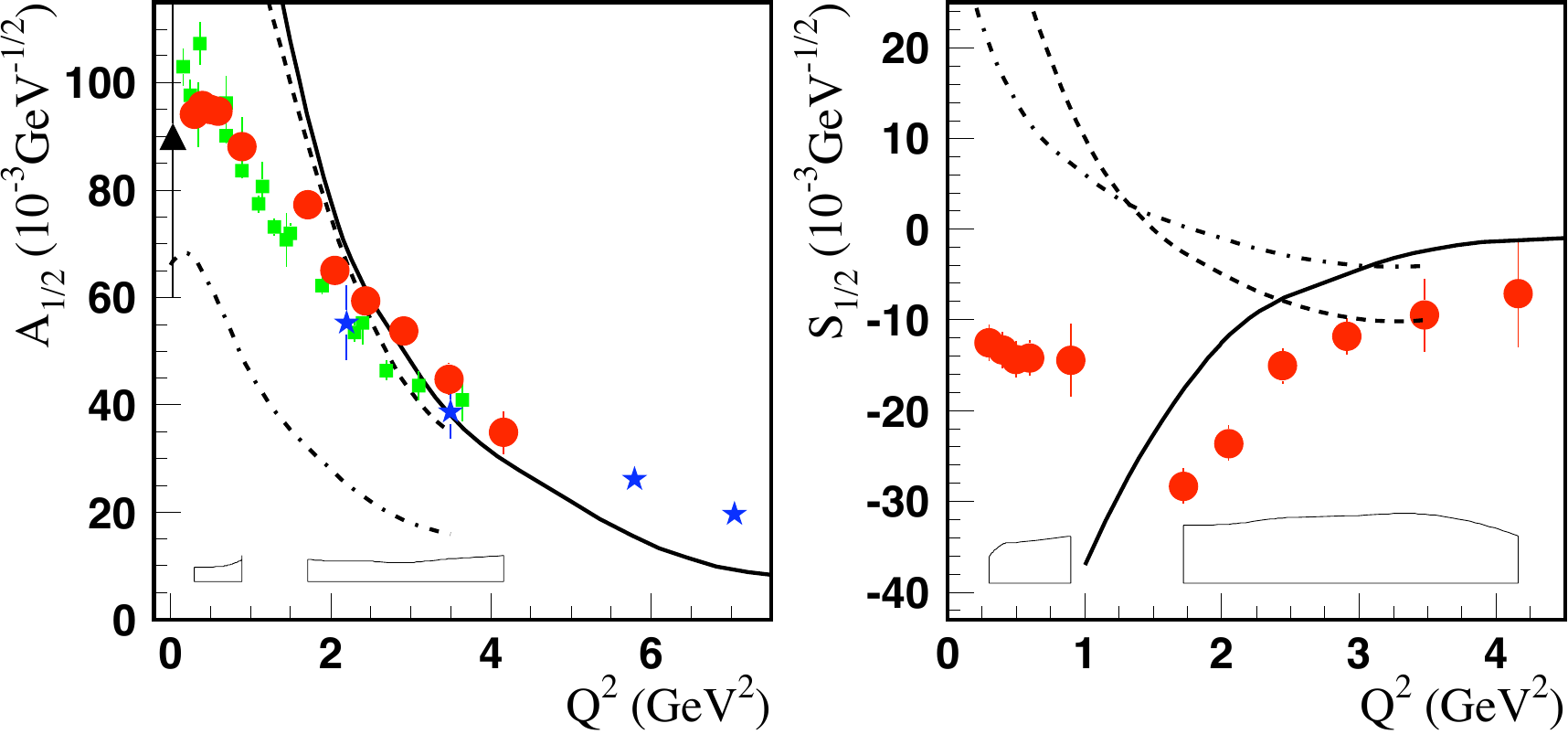} 
\caption {\small The transition amplitude $A_{1/2}$ (left) for the $S_{11}(1535)$. The full circles are 
from the analysis of the CLAS $n\pi^+$ and $p\pi^\circ$ data\protect\cite{aznauryan-1,aznauryan-2}. 
The other data are from the analysis of $p\eta$ data~\protect\cite{thompson,denizli,armstrong,dalton}. 
The curves represent constituent quark model calculations of~\cite{capstick} (dashed), 
~\cite{salme} (dashed-dotted), and~\cite{braun} (solid).}
\label{s11}
\end{figure}
\section{The $N(1535)S_{11}$ state} 
This state has been studied extensively in the $p\eta$ channel, where it 
appears as an isolated resonance near the $N\eta$ threshold.
Phenomenological analyses of data from CLAS~\cite{thompson,denizli} and Hall C~\cite{armstrong,dalton} 
have resulted in the $Q^2$ evolution of the transverse transition 
amplitude $A_{1/2}$ from $\eta$ electroproduction data. 
However, there are two remaining important uncertainties that need to be examined. 
The first uncertainty is due to the 
branching ratio of the coupling $S_{11}(1535) \rightarrow p\eta$. The PDG~\cite{pdg2008} 
gives ranges of  $\beta^{PDG}_{N\eta} = 0.45 - 0.60$ and  $\beta^{PDG}_{N\pi} = 0.35 - 0.55$, 
which adds a large uncertainty to the resulting helicity amplitudes. 
Since this state has very small coupling to channels other than $N\eta$ and $N\pi$, 
a measurement of the reaction $ep\rightarrow e\pi^+n$ can reduce this uncertainty.
Adjusting $\beta_{N\pi} = 0.48$ and  $\beta_{N\eta}=0.46$ brings the two data sets into 
excellent agreement, as shown in Fig.~\ref{s11}. 
The second uncertainty comes from the lack of precise information on the longitudinal coupling. 
This contribution is usually neglected when analyzing the $p\eta$ channel. 
An important advantage of the $N\pi$ channel is that it is also 
sensitive to the longitudinal transition amplitude $S_{1/2}$ resulting from a significant $s-p$ wave
interference with the nearby $p$-wave amplitude of the $P_{11}(1440)$. Since the  $P_{11}(1440)$
does not couple to $p\eta$, this channel has very little sensitivity to the $S_{1/2}$ amplitude.   
 
\section{Helicity structure of the $D_{13}(1520)$ }

A longstanding prediction~\cite{close} of the dynamical constituent quark model is the helicity 
switch from the dominance of the $A_{3/2}$ amplitude at the photon point to $A_{1/2}$ dominance 
at $Q^2 > 1$~GeV$^2$. Indications of such behavior have been observed in previous 
analysis~\cite{burkert-lee,maid07}, but analyses have been hampered by incomplete kinematical 
coverage of data, and the scarceness of $n\pi^+$ data, which are most sensitive to the excitation 
of the state. The new CLAS data have largely eliminated this shortcoming.   
\begin{figure}[h]
\includegraphics[width=14cm, height=5.5cm]{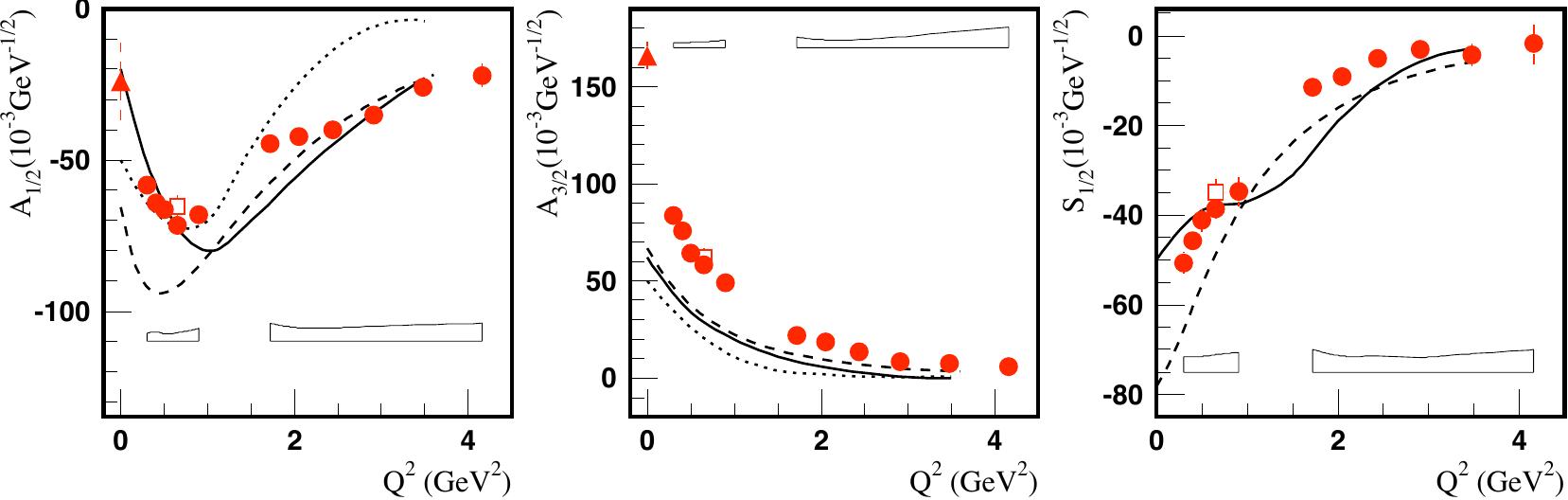}
\caption{Electrocoupling amplitudes $A_{1/2}$ (left) and $A_{3/2}$ 
(middle) and $S_{1/2}$ (right) for the $D_{13}(1520)$. Model curves as in Fig.2.} 
\label{d13}
\end{figure} 
Figure~\ref{d13}
shows the CLAS results for the electrocouplings. The $A_{3/2}$ amplitude is large at the real 
photon point and decreasing rapidly 
in strength with increasing $Q^2$. $A_{1/2}$ is small at the photon point and increases 
rapidly in magnitude with increasing $Q^2$. At high $Q^2$ $A_{1/2}$ becomes the 
dominant amplitude, which confirms the early prediction of the constituent quark model. 
\section{Conclusions}
The past decade of experimental research on the electroproduction of 
pseudoscalar mesons, has led to consistent set of data on transition amplitudes for 
several of the lower mass excited states of the nucleon. This set of the most precise 
amplitudes to date allows us to put to test models of the nucleon 
structure in terms of the effective degrees of freedom.

\vspace{0.3cm}   
This work was performed under DOE contract DE-AC05-060R23177.

\end{document}